\documentclass[conference]{IEEEtran}
\IEEEoverridecommandlockouts
\usepackage{cite}
\usepackage{amsmath,amssymb,amsfonts}
\usepackage{algorithmic}
\usepackage{graphicx}
\usepackage{textcomp}
\usepackage{xcolor}
\def\BibTeX{{\rm B\kern-.05em{\sc i\kern-.025em b}\kern-.08em
    T\kern-.1667em\lower.7ex\hbox{E}\kern-.125emX}}
\makeatletter 
\newcommand{\linebreakand}{%
  \end{@IEEEauthorhalign}
  \hfill\mbox{}\par
  \mbox{}\hfill\begin{@IEEEauthorhalign}
}
\makeatother 
\begin{document}

\title{Predicting ejection fraction from chest x-rays using computer vision for diagnosing heart failure\\
}

\author{\IEEEauthorblockN{Walt Williams}
\IEEEauthorblockA{\textit{Institute for Applied Computational Science} \\
\textit{Harvard University}\\
Boston, United States \\
wwilliams@g.harvard.edu}
\and
\IEEEauthorblockN{Yanran Li}
\IEEEauthorblockA{\textit{Department of Biostatistics} \\
\textit{Harvard University}\\
Boston, United States \\
yanranli@hsph.harvard.edu}
\and
\IEEEauthorblockN{Rohan Doshi}
\IEEEauthorblockA{\textit{Department of Engineering Sciences} \\
\textit{Harvard University}\\
Boston, United States \\
rohandoshi@g.harvard.edu
}
\linebreakand 
\IEEEauthorblockN{Kexuan Liang}
\IEEEauthorblockA{\textit{Department of Biostatistics} \\
\textit{Harvard University}\\
Boston, United States \\
kexuanliang@hsph.harvard.edu}
 }

\maketitle

\begin{abstract}
Heart failure remains a major public health challenge with growing costs. Ejection fraction (EF) is a key metric for diagnosis and management of heart failure; however estimation of EF using echocardiography remains expensive for the healthcare system and subject to intra/inter operator variability.   While chest x-ray (CXR) are quick, inexpensive, and require less expertise, they do not provide sufficient information to the human eye to estimate EF. This work explores the efficacy of computer vision techniques to predict reduced EFsolely from CXRs. We studied a dataset of 3488 CXRs from the MIMIC CXR-jpg (MCR) dataset. Our work established binary classification benchmarks using multiple state-of-the-art convolutional neural network architectures. The subsequent analysis shows increasing model sizes from 8M to 23M parameters improved classification performance without overfitting the dataset. We further show how data augmentation techniques such as CXR rotation and random cropped resizing further improved model performance another ~5\%. Finally, we conduct an error analysis using saliency maps and Grad-CAM maps to better understand the failure models of convolutional models on this task.

\end{abstract}


\section{Introduction}

Heart failure (HF) is a prevalent chronic  condition affecting between 1-2\% of the adult population. Heart Failure is classified into two subgroups:  heart failure with reduced ejection fraction (HFrEF) and heart failure with preserved ejection fraction (HFpEF). Echocardiograms are used to measure left ventricular ejection fraction (LVEF). When a patient has HF with LVEF higher than 50\%, a diagnosis of HFpEF is made; and when LVEF is lower than 40\%, a diagnosis of HFrEF is made. This distinction is critical as HFpEF and HFrEF are managed differently. 

Despite recent advances, 5-year mortality from heart failure remains high, at approximately 50\% \cite{b1}. Given its high incidence and mortality, the American Heart Association recommends a set of Guideline Directed Medical Therapies (GDMT), which encompass clinical evaluation, diagnostic testing, and pharmacological and procedural treatments \cite{b2}. Diagnosis often includes, but is not limited to, transthoracic echocardiograms (TTEs), which evaluate heart function. TTEs are currently the accepted standard for determining LVEF. Early diagnosis is critical, as patients not receiving Guideline Directed Medical Therapies (GDMT) for HFrEF have a 37\% increased two-year mortality\cite{b22}. 

However, LVEF estimations from TTEs are subject to human interpretation and variation; intraobserver and interobserver variability of standard echocardiographic left ventricular ejection fraction (EF) assessment is reported to be 8–21\% and 6–13\% \cite{b3}. Additionally, TTEs cost the healthcare system hundreds of dollars (\$200-400). 

The development of an inexpensive, standardized algorithm to estimate LVEF would alleviate the time and expertise burden required to perform and interpret TTEs, the cost to the healthcare system, and the variability of LVEF assessments. One solution is estimating LVEF using machine learning models trained on chest x-rays (CXR), which are quicker and cheaper to perform since each CXR costs approximately \$90 cost to the healthcare system and requires less expertise.

\section{Related Works}
Previous work has shown that machine learning models can be used to predict a patient’s LVEF from a CXR. Related work in this field includes the use of AI for the diagnosis of heart failure from medical images. For example, in a study by Matsumoto et al.\cite{b17}, a deep learning algorithm was used to diagnose heart failure from chest X-ray images, achieving high accuracy and performance. Another study by Que et al.\cite{b18} used a deep learning algorithm to automatically detect cardiomegaly, a common symptom of heart failure, from chest X-ray images. In addition to the use of AI in medical imaging, there is also a growing body of work on the use of AI for the diagnosis and treatment of heart failure. For example, a study by Ambrosy et al.\cite{b19} found that hospitalizations for heart failure are a significant global health and economic burden, highlighting the need for more effective diagnostic and treatment methods. Another study by Kilic et al.\cite{b20} reviewed the use of AI and machine learning in cardiovascular healthcare, highlighting the potential of these technologies in improving the accuracy and reliability of diagnostic methods. Overall, the use of AI in medical imaging, and specifically for the detection and classification of heart failure, has shown great potential in improving the accuracy and reliability of diagnostic methods. 

Besides, the study by Chih-Wei Hsiang et al.\cite{b21} used a large dataset of 90,000 chest X-ray (CXR) images from an academic medical center to develop an artificial intelligence (AI) model for the detection of left ventricular systolic dysfunction (LVSD). The study used a binary classification approach to predict reduced left ventricular ejection fraction (LVEF) defined as $\leq 35\%$. However, few studies in this area focuses on the architectures to establish binary classification benchmarks for LVSD detection or investigate the impact of model size and data augmentation techniques on classification performance. Moreover, we haven't found those works conduct error analysis using saliency maps and Grad-CAM maps to visualize and interpret the model's decision-making process. Therefore, we try to contribute to the growing body of work on the use of AI for the diagnosis and treatment of heart failure, highlighting the potential of deep learning methods for improving the accuracy and interpretability of diagnostic models.

\section{Methodology}

\subsection{Dataset Collection}
MIMIC CXR-jpg (MCR) \cite{b6, b7, b8} dataset is a large publicly available dataset of 377,110 chest CXRs converted to jpg format. The images were collected from Electronic Health Records from the Beth Israel Deaconess Medical Center (BIDMC) emergency room from 2011-2016. For our study, we were only interested in images diagnosed with HFpEF or HRrEF, so we filtered the dataset into only those CXRs with the desired ICD codes using MIMIC IV \cite{b9}. Our patient cohort consists of 3,488 CXRs, with 2,010 CXRs diagnosed with HFrEF and the remaining 1,478 CXRs diagnosed with HFpEF. The median age for patients in our dataset was 71, with an interquartile range of 61-81. The gender distribution of patients was 1,579 (45.3\%) female and 1,909 (54.7\%) male patients. And the distribution of races and ethnicities can be found in Table \ref{tab1}. To prepare our data, we first normalized each image pixel's value to between 0 and 1 by dividing RGB values by 255. We then split the data into a training set, testing set, and validation set using a 65\%-25\%-10\% split. Finally, we ensured no data leakage before continuing with training. 

\begin{table}[htbp]
\caption{Summary of race/ethnicity of our patient cohort}
\begin{center}
\begin{tabular}{|c|c|}
\hline
\textbf{Race / Ethnicity} & \textbf{Number in data (\% of data)} \\
\hline
\text{American Indian/Alaska Native} & \text{22 (0.6\%)} \\
\hline
\text{Asian} & \text{100 (3\%)} \\
\hline
\text{Black} & \text{680 (19.4\%)} \\
\hline
\text{Hispanic/ Latino} & \text{187 (5.4\%)} \\
\hline
\text{Other} & \text{169 (4.8\%)} \\
\hline
\text{Unknown} & \text{144 (4.1\%)} \\
\hline
\text{Unable to Obtain} & \text{20 (0.6\%)} \\
\hline
\text{Multiple Race/Ethnicity} & \text{11 (0.3\%)} \\
\hline
\text{Declined to answer} & \text{41 (1.2\%)} \\
\hline
\text{White/Caucasian} & \text{2095 (60\%)} \\
\hline
\end{tabular}
\label{tab1}
\end{center}
\end{table}

\subsection{Model Comparison}
We evaluated multiple deep learning neural network architectures from the the computer vision literature to study how models with varying complexity perform on our CXR dataset. The architectures used were ResNet50 \cite{b10}, EfficientNet (b0) \cite{b11}, and DenseNet121 \cite{b12}. All models were trained using Adaptive Moment Estimation (Adam) as the optimizer with an initial learning rate of .001. The learning rate was divided by ten after every five epochs of training when there was no performance improvement. 


\subsection{Model Augmentation}
We observed minor variances in the presentation of the input x-ray images fed into the model, which correlates with model misclassification on slightly permuted inputs. So, we apply two image transformations during training to improve model robustness to variance in X-ray inputs. First, we randomly rotated the input images by $\pm10$ degrees to simulate how input x-rays and their underlying human subjects may not be vertically aligned. Second, we randomly cropped and resized the inputs, varying the scale from 0.75x to 1.00x the original image size. This serves to emulate how input x-rays may have cropping at the edges of the image.

\subsection{Error Analysis and Model Interpretation}
We implemented two gradient-based methods to visualize and give insight to the model's decision making. this helps us better analyze the misclassified cases' using heatmaps. For the first method, saliency maps (Vanilla Gradient method) \cite{b13}, we calculate and plot the magnitude of the gradients of the loss value for the interested class with respect to the input pixels. For the second method, Grad-CAM (Gradient-weighted Class Activation Map) \cite{b14}, the gradient is usually back-propagated to the last convolutional layer to produce a coarse localization map that highlights important regions of the image. 

\section{Results}
\subsection{Model Performance}
Of the total 872 samples, 600 images were correctly classified, and 272 images were misclassified; 439 images were classifified with high confidence ($probability>0.9$) and 151 images were with low confidence($probability<0.1$). Among the 272 misclassified images, 118 were with high confidence and 32 were with low confidence. We used the saliency map and Grad-CAM to generate heatmaps for the samples (6 examples in each group).

One of the open questions is whether these convolutional models are overparameterized and overfitting the training data. By comparing the parameter counts, and hence the complexity, of the three convolutional models, we see how the model performance increases as we increase model parameterization. In Table II, as we increase parameter count from 8M with DenseNet 121 to 11M with EfficientNet B0 and then 23M with ResNet50, we see a steady improvement in precision, recall, and F1-score for both the reduced EF and preserved EF classes. This suggests that these convolutional models are not over-parameterized enough to overfit the training data.

\begin{table}[htbp]
\caption{A comparison of our model performance with various architectures.}
\begin{center}
\begin{tabular}{|c|c|c|c|c|}
\hline
\textbf{Architecture} & \textbf{Class} & \textbf{Precision} & \textbf{Recall} & \textbf{F1-score} \\
\hline
\text{Densenet 121} & \text{Reduced EF} & \text{0.66} & \text{0.29} & \text{0.41} \\
\cline{2-5} 
\text{(8M params)} & \text{Preserved EF} & \text{0.62} & \text{0.89} & \text{0.73} \\
\hline

\text{EfficientNet B0} & \text{Reduced EF} & \text{0.70} & \text{0.52} & \text{0.60} \\
\cline{2-5} 
\text{(11M params)} & \text{Preserved EF} & \text{0.69} & \text{0.83} & \text{0.76} \\
\hline
\text{ResNet50} & \text{Reduced EF} & \text{0.63} & \text{0.58} & \text{0.60} \\
\cline{2-5} 
\text{(23M params)} & \text{Preserved EF} & \text{0.68} & \text{0.73} & \text{0.71} \\
\hline
\end{tabular}
\label{tab2}
\end{center}
\end{table}

\subsection{Data Augmentation}
Table \ref{tab3} shows how data augmentation using basic image transformations can further improve model performance. After we add data augmentation to ResNet50, our best model, we see a 5\% improvements for Precision, Recall, and F1-score, across both the reduced EF and preserved EF classes. This suggests that data augmentations during training improves model robustness to input permutations in the test set.

\begin{table}[htbp]
\caption{The model performance after applying data augmentation.}
\begin{center}
\begin{tabular}{|c|c|c|c|c|}
\hline
\textbf{Architecture} & \textbf{Class} & \textbf{Precision} & \textbf{Recall} & \textbf{F1-score} \\
\hline
\text{ResNet50} & \text{Reduced EF} & \text{0.63} & \text{0.58} & \text{0.60} \\
\cline{2-5} 
\text{(Baseline)} & \text{Preserved EF} &  \text{0.68} & \text{0.73} & \text{0.71} \\
\hline
\text{ResNet50} & \text{Reduced EF} & \text{0.66} & \text{0.58} & \text{0.62} \\
\cline{2-5} 
\text{(with data augmentation} & \text{Preserved EF} & \text{0.71} & \text{0.77} & \text{0.75} \\
\hline
\end{tabular}
\label{tab3}
\end{center}
\end{table}

\subsection{Error Analysis and Model Interpretation}
\begin{figure}[htbp]
\centerline{\includegraphics[width=80mm]{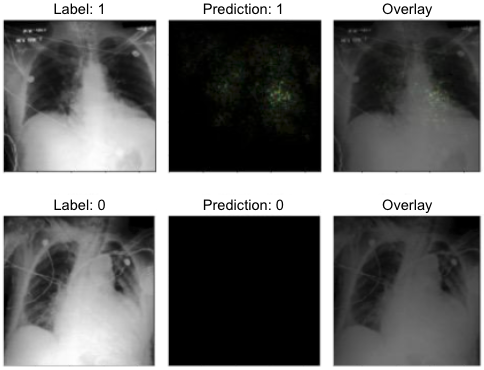}}
\caption{Saliency maps for the correct classified cases.}
\label{fig1}
\end{figure}

\begin{figure}[htbp]
\centerline{\includegraphics[width=80mm]{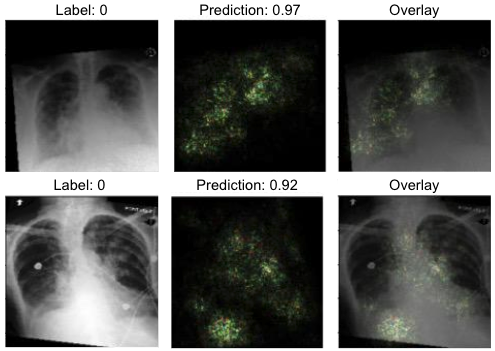}}
\caption{Saliency maps for the false positive cases.}
\label{fig2}
\end{figure}

\begin{figure}[htbp]
\centerline{\includegraphics[width=80mm]{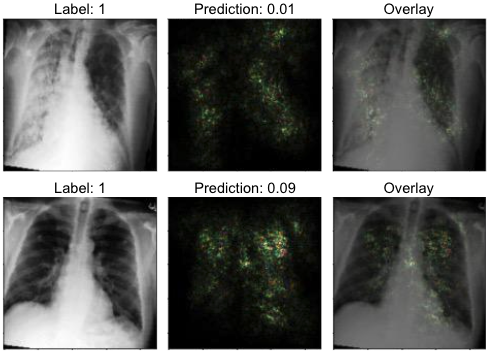}}
\caption{Saliency maps for the false negative cases.}
\label{fig3}
\end{figure}

Figure \ref{fig1}, \ref{fig2}, \ref{fig3} show the saliency maps for correctly classified, false positive, and false negative cases produced by the baseline ResNet50 model. The left panel shows the original chest images of the patients. The middle panel shows the saliency maps. And the right panel shows the saliency maps overlaid on top of the chest images. The brigtness of the color indicates which pixels had the most significant influence on the model's final prediction.

\section{Discussion}

\subsection{Limitations}
Supervised machine learning requires a large amount of accurate and diverse training data to be robust against overfitting and bias. The size of our patient cohort, and hence dataset, is limited. Furthermore, the race distribution in our cohort is skewed. Within our population, white patients outnumbered those from Asians, Black, and Hispanics backgrounds. The race imbalance in the data could lead to a biased machine-learning model and uneven diagnosis outcomes \cite{b15, b16}. We see this in Table \ref{tab4}, which shows the performance  of our best model on a holdout test set broken down by race and gender. 

\begin{table}[htbp]
\caption{A comparison of ResNet50 performance across racial groups}
\begin{center}
\begin{tabular}{|c|c|c|c|c|}
\hline
\textbf{Race/Gender} & \textbf{Class} & \textbf{Precision} & \textbf{Recall} & \textbf{F1-score} \\
\hline
\text{White} & \text{Reduced EF} & \text{0.64} & \text{0.62} & \text{0.63} \\
\cline{2-5} 
\text{} & \text{Preserved EF} & \text{0.72} & \text{0.74} & \text{0.73} \\
\hline
\text{Black} & \text{Reduced EF} & \text{0.38} & \text{0.32} & \text{0.35} \\
\cline{2-5} 
\text{} & \text{Preserved EF} & \text{0.56} & \text{0.62} & \text{0.59} \\
\hline
\text{Hispanic/Latino} & \text{Reduced EF} & \text{0.57} & \text{0.86} & \text{0.69} \\
\cline{2-5} 
\text{} & \text{Preserved EF} & \text{0.89} & \text{0.64} & \text{0.74} \\

\hline
\text{Asian} & \text{Reduced EF} & \text{0.90} & \text{0.64} & \text{0.75} \\
\cline{2-5} 
\text{} & \text{Preserved EF} & \text{0.69} & \text{0.92} & \text{0.79} \\
\hline
\text{Male} & \text{Reduced EF} & \text{0.71} & \text{0.68} & \text{0.69} \\
\cline{2-5} 
\text{} & \text{Preserved EF} & \text{0.64} & \text{0.68} & \text{0.66} \\\hline
\text{Female} & \text{Reduced EF} & \text{0.52} & \text{0.45} & \text{0.48} \\
\cline{2-5} 
\text{} & \text{Preserved EF} & \text{0.78} & \text{0.82} & \text{0.80} \\
\hline
\end{tabular}
\label{tab4}
\end{center}
\end{table}

Another limitation of our study is the reliance on ICD-10 codes for identifying HFpEF and HFrEF. Previous studies suggest that ICD heart failure subtype codes have weaker accuracy compared to echocardiograms in predicting the LVEF values from electronic medical record review. The ICD HF subtype choice should have reasonable accuracy in the MIMIC database. However, if the ICD HF subtype code is inaccurate or not up to date with the corresponding CXR, this would cause mislabeling in our model’s training set.

\subsection{Future Work}
First, we hope to improve visualization used for model explainabiltiy. When we generated saliency maps, we implemented a backpropagation algorithm to compute the gradients of the logits with respect to the network's input. However, the resulting heatmaps could not thoroughly explain the model. It is possible to use deconvolutional networks or guided backpropagation algorithms to provide better heatmaps. For example, combining Grad-CAM with guided backpropagation might provide higher resolution. This guided Grad-CAM approach could be highly class-discriminative and help our error analysis.

Additionally, we want to explore better pre-training approaches to improve model performance and reduce domain mismatch. Our baseline ResNet model loaded in ImageNet weights, which allowed for good baseline feature extraction and quick model convergence. But this choice introduced domain mismatch risk between the ImageNet domain (internet images) and CXRs. In the future, we'd like to explore using pretrained weights learned from a medical dataset such as CheXpert. This could better guide the model to capture pulmonary-related features.

In future work, we will also include higher-quality and larger datasets to create more performant and robust models ready for clinical use. Real-world application will need extensive clinical expertise to deeply evaluate these models and integrate them into existing clinical workflows.

\section{Conclusion}
In summary, this work further studies how computer vision techniques can be applied to heart failure 
CXRs to classify LVEF, offering a low lost alternative to transthoracic echocardiograms. Our best model shows good performance with a misclassification rate 31.2\%. This performance level could serve as a starting point for further research and refinement, but further improvements are needed to reach a performance level for clinical deployment. Among our contributions, we show how improving model size can increase model performance and that data augmentation can improve model robustness. Furthermore, we demonstrate how two complementary visualization approaches, the saliency map and Grad-CAM, can improve model explainability by showing which input pixels from the CXR most informed the model's ultimate prediction.




\end{document}